\documentstyle[12pt,epsf]{article}

\catcode`@=11
\def\chkspace{%
  \relax   % Calm down any expanding \if's.
  \begingroup\ifhmode\aftergroup\dochksp@ce\fi\endgroup}
\def\dochksp@ce{%
  \unskip              % Remove any preceding horizontal glue
  \futurelet\chkspct@k\d@chkspc  % Grab the next token and look ahead
}
\def\d@chkspc{%
  \let\nxtsp@ce=\relax
  \ifx\chkspct@k.\else     % Don't put spaces before punctuation ...
    \ifx\chkspct@k,\else
      \ifx\chkspct@k;\else
        \ifx\chkspct@k!\else
          \ifx\chkspct@k?\else
            \ifx\chkspct@k:\else
              \ifx\chkspct@k)\else
              \ifx\chkspct@k(\else
                \ifx\chkspct@k]\else
                  \ifx\chkspct@k-\else
                    \ifx\chkspct@k\egroup\else  % or close groups.
                      \let\nxtsp@ce=\put@space  % Put a space everywhere else.
                    \fi
                  \fi
                \fi
              \fi
              \fi
            \fi
          \fi
        \fi
      \fi
    \fi
  \fi
  \nxtsp@ce
}
\def\put@space{$\;$}
\catcode`@=12

\def\bx{\bar x}

\def\etal{{\it et al.}\chkspace}

\def\ep{{e$^+$e$^-$}\chkspace}

\def\gluino{\relax\ifmmode \tilde{g} \else $\tilde{g}$ \fi\chkspace}

\def\bbar{$\overline{\rm b}$\chkspace}

\def\bbrm{\relax\ifmmode {\rm b}\bar{\rm b}
       \else ${\rm b}\bar{\rm b}$ \fi\chkspace}
\def\bb{$b\bar{b}$ \chkspace}
\def\ccrm{\relax\ifmmode {\rm c}\bar{\rm c}
       \else ${\rm c}\bar{\rm c}$ \fi\chkspace}
\def\cc{$c\bar{c}$ \chkspace}
\def\tt{\relax\ifmmode {\rm t}\bar{\rm t}
       \else ${\rm t}\bar{\rm t}$ \fi\chkspace}
\def\ss{\relax\ifmmode {\rm s}\bar{\rm s}
       \else ${\rm s}\bar{\rm s}$ \fi\chkspace}
\def\uu{\relax\ifmmode {\rm u}\bar{\rm u}
       \else ${\rm u}\bar{\rm u}$ \fi\chkspace}
\def\dd{\relax\ifmmode {\rm d}\bar{\rm d}
       \else ${\rm d}\bar{\rm d}$ \fi\chkspace}

\def\qqg{\relax\ifmmode {\rm q}\overline{\rm q}{\rm g}
\else q$\overline{\rm q}$g \fi\chkspace}

\def\bbg{$b\overline{b}g$\chkspace}
\def\ccg{$c\overline{c}g$\chkspace}

\def\afb{\relax\ifmmode A_{FB} \else
{{$A_{FB}$}}\fi\chkspace}
\def\afbb{\relax\ifmmode A_{FB}^b \else
{{$A_{FB}^b$}}\fi\chkspace}
\def\pafb{\relax\ifmmode \tilde{A}_{FB} \else
{{$\tilde{A}_{FB}$}}\fi\chkspace}
\def\pafbb{\relax\ifmmode \tilde{A}_{FB}^b \else
{{$\tilde{A}_{FB}^b$}}\fi\chkspace}

\def\pafbzo{\relax\ifmmode \tilde{A}_{FB}|_{O(0)} \else
{{$\tilde{A}_{FB}|_{O(0)}$}}\fi\chkspace}
\def\pafbfo{\relax\ifmmode \tilde{A}_{FB}|_{\oalp} \else
{{$\tilde{A}_{FB}|_{\oalp}$}}\fi\chkspace}
\def\pafbso{\relax\ifmmode \tilde{A}_{FB}|_{\oalpsq} \else
{{$\tilde{A}_{FB}|_{\oalpsq}$}}\fi\chkspace}
\def\pafbto{\relax\ifmmode \tilde{A}_{FB}|_{\oalpc} \else
{{$\tilde{A}_{FB}|_{\oalpc}$}}\fi\chkspace}

\def\pafbbzo{\relax\ifmmode \tilde{A}_{FB}^b|_{O(0)} \else
{{$\tilde{A}_{FB}^b|_{O(0)}$}}\fi\chkspace}
\def\pafbbfo{\relax\ifmmode \tilde{A}_{FB}^b|_{\oalp} \else
{{$\tilde{A}_{FB}^b|_{\oalp}$}}\fi\chkspace}
\def\pafbbso{\relax\ifmmode \tilde{A}_{FB}^b|_{\oalpsq} \else
{{$\tilde{A}_{FB}^b|_{\oalpsq}$}}\fi\chkspace}
\def\pafbbto{\relax\ifmmode \tilde{A}_{FB}^b|_{\oalpc} \else
{{$\tilde{A}_{FB}^b|_{\oalpc}$}}\fi\chkspace}

\def\afbo0{\tilde{A}_{FB}|_{O(0)}}
\def\afbo1{\tilde{A}_{FB}|_{\oalp}}
\def\afbo2{\tilde{A}_{FB}|_{\oalpsq}}
\def\afbo3{\tilde{A}_{FB}|_{\oalpc}}

\def\lam{\relax\ifmmode \Lambda_{\overline{MS}}
       \else {{$\Lambda_{\overline{MS}}$}}\fi\chkspace}
\def\lamuds{\relax\ifmmode \Lambda^{(3)}_{\overline{MS}}
       \else {{$\Lambda^{(3)}_{\overline{MS}}$}}\fi\chkspace}
\def\lamudsc{\relax\ifmmode \Lambda^{(4)}_{\overline{MS}}
       \else $\Lambda^{(4)}_{\overline{MS}}$\fi\chkspace}
\def\lamudscb{\relax\ifmmode \Lambda^{(5)}_{\overline{MS}}
       \else $\Lambda^{(5)}_{\overline{MS}}$\fi\chkspace}

\def\alp{\relax\ifmmode \alpha_s\else $\alpha_s$\fi\chkspace}
\def\alpbar{\relax\ifmmode \bar{\alpha_s}
       \else $\bar{\alpha_s}$\fi\chkspace}
\def\alpmz{\relax\ifmmode \alpha_s(M_Z)\else $\alpha_s(M_Z)$\fi\chkspace}
\def\alpmzsq{\relax\ifmmode \alpha_s(M_Z^2)
       \else $\alpha_s(M_Z^2)$\fi\chkspace}

\def\oalp{\relax\ifmmode O(\alpha_s)\else{{O($\alpha_s$)}}\fi\chkspace}
\def\oalpsq{\relax\ifmmode O(\alpha_s^2)
           \else{{O($\alpha_s^2$)}}\fi\chkspace}
\def\oalpc{\relax\ifmmode O(\alpha_s^3)
           \else{{O($\alpha_s^3$)}}\fi\chkspace}
\def\oalpf{\relax\ifmmode O(\alpha_s^4)
           \else{{O($\alpha_s^4$)}}\fi\chkspace}

\def\rb{\relax\ifmmode R_3^b/R_3^{all}
           \else{{$R_3^b/R_3^{all}$}}\fi\chkspace}
\def\rc{\relax\ifmmode R_3^c/R_3^{all}
           \else{{$R_3^c/R_3^{all}$}}\fi\chkspace}
\def\ruds{\relax\ifmmode R_3^{uds}/R_3^{all}
           \else{{$R_3^{uds}/R_3^{all}$}}\fi\chkspace}
\def\ri{\relax\ifmmode R_3^i/R_3^{all}
           \else{{$R_3^i/R_3^{all}$}}\fi\chkspace}
\def\rj{\relax\ifmmode R_3^j/R_3^{all}
           \else{{$R_3^j/R_3^{all}$}}\fi\chkspace}
\def\alpi{\relax\ifmmode \alpha^i_s/\alpha^{all}_s
           \else{{$\alpha^i_s/\alpha^{all}_s$}}\fi\chkspace}

\def\z0{{$Z^0$}\chkspace}
\def\Dst{\relax\ifmmode {\rm D}^* \else {D$^*$}\fi\chkspace}
\def\Dpl{\relax\ifmmode {\rm D}^+ \else {D$^+$}\fi\chkspace}
\def\D0{\relax\ifmmode {\rm D}^0 \else {D$^0$}\fi\chkspace}
\def\Kst{\relax\ifmmode {\rm K}^* \else {K$^*$}\fi\chkspace}
\def\K0{\relax\ifmmode {\rm K}^0_s \else {K$^0_s$}\fi\chkspace}
\def\Kpl{\relax\ifmmode {\rm K}^+ \else {K$^+$}\fi\chkspace}
\def\Kstz{\relax\ifmmode {\rm K}^{*0} \else {K$^{*0}$}\fi\chkspace}

\renewcommand{\baselinestretch}{1.5}

 1

\catcode`\@=11 % This allows us to modify PLAIN macros.
\makeatletter
\def\@seccntformat#1{\csname the#1\endcsname.\hskip 1em}
\makeatother
\begin{document}
\thispagestyle{empty}
\begin{flushright}
{\footnotesize\renewcommand{\baselinestretch}{.75}
  SLAC--PUB--8156\\
June 1999\\
}
\end{flushright}

\vskip 1truecm
 
\begin{center}
 {\Large \bf Symmetry Tests in Polarized \z0 Decays to b${\bar {\bf b}}$g$^*$}
 
\vspace {1.0cm}
 
 {\bf The SLD Collaboration$^{**}$}\\
Stanford Linear Accelerator Center \\
Stanford University, Stanford, CA~94309
 
\vspace{1.5cm}
 
\end{center}
 
\normalsize
%===============
% Abstract
%===============
 
%\hspace{1in}
 
\begin{center}
{\bf ABSTRACT }
\end{center}

{\small
\noindent
 Angular asymmetries have been measured in polarized \z0 decays to  
 b${\bar {\rm b}}$g  collected by the SLD experiment at the SLC.
 A high purity b${\bar {\rm b}}$g event sample is selected by utilizing 
 $B$ lifetime information given by the SLD CCD pixel
 vertex detector and the stable micron-size SLC beams, 
 and the b- and ${\bar {\rm b}}$-jets are identified using  
 lifetime information and momentum-weighted track charge.
 The forward-backward asymmetry is observed 
  in the b-quark polar angle distribution, and the parity-violation parameter
 is measured  to test the Standard Model. 
 Two angular correlations between the three-jet plane and the \z0 polarization
 are studied. 
 The CP-even and T-odd, and the CP-odd and T-odd, 
 angular asymmetries
 are sensitive to physics beyond the Standard Model.
 The latter requires tagging both the b- and  ${\bar {\rm b}}$-jet. 
 We measure the 
 expectation values of these quantities to be consistent with zero and set 
 limits on the correlations at the 5\% level.

}
 
\vfil
 
\noindent
Contributed to: the International Europhysics Conference on High Energy Physics,
15-21 July 1999, Tampere, Finland; Ref. 1\_183, and to the XIXth International 
Symposium on Lepton and Photon Interactions, August 9-14 1999, Stanford, USA.

\vskip .3truecm

\noindent
$^*$Work supported by Department of Energy contract DE-AC03-76SF00515 (SLAC).

\eject
  
\section{Introduction}

 The forward-backward polar-angle asymmetry in  hadronic \z0 decays to two jets
has been investigated extensively at SLC and LEP to test the predictions of
the electroweak theory of parity-violation in the $Z^0q{\bar q}$ coupling. 
In particular, at SLC where the
electron beam is highly polarized, the left-right-forward-backward 
asymmetry removes the dependence on the $Z^0e^+e^-$ coupling and
is directly sensitive to the $Z^0q{\bar q}$ coupling. 
The experimental results are found to be consistent with the 
theory to within  experimental uncertainties of a few percent~\cite{ew}.
Hadronic  \z0 decays to three jets can be interpreted in 
terms of the fundamental process \z0$\rightarrow$ \qqg where one of
the quarks has radiated a gluon.
Given the success of the electroweak theory in predicting the two-jet
polar-angle asymmetry, similar angular asymmetries can be measured in
three-jet events to test Quantum Chromodynamics (QCD).
 The \z0$\rightarrow$\bbg final state is
particularly interesting as a search-ground for possible new physics processes
beyond the Standard Model, and 
a high purity sample can be obtained with high
efficiency due to the large mass and  long lifetime of B-hadrons.
Here we report the first experimental study of angular asymmetries in 
polarized \z0 decays to \bbg.
\section{Angular asymmetries in \z0 $\rightarrow$ \qqg}

The differential cross section for \ep $\rightarrow$
\qqg can be expressed as
~\cite{qqg}
$$2\pi{{d^4\sigma}\over{d(\cos\theta)d\chi dxd\bx}} = $$
$$[~{3\over 8}(1+\cos^2\theta){{d^2\sigma_U}\over{dxd\bx}}+
{3\over4}\sin^2\theta{{d^2\sigma_L}\over{dxd\bx}}   \\
+{3\over4}\sin^2\theta\cos 2\chi {{d^2\sigma_T}\over{dxd\bx}}
+{3\over{2\sqrt{2}}}\sin 2\theta\cos\chi {{d^2\sigma_I}\over{dxd\bx}}~
]~h_f^{(1)}(s) $$
\begin{equation}
+~[~{3\over4}\cos\theta {{d^2\sigma_P}\over{dxd\bx}}
-{3\over\sqrt{2}}\sin\theta\cos\chi {{d^2\sigma_A}\over{dxd\bx}}~]~
h_f^{(2)}(s),
\end{equation}
where $x$ and $\bar x$ are the scaled momenta of the quark and anti-quark,
respectively, 
$\theta$ is the polar angle of the thrust axis~\cite{thrust} 
w.r.t. the electron beam,
and $\chi$ is the azimuthal angle of the event plane w.r.t. the quark-electron
plane. Here the thrust axis is defined so that it is parallel to the quark
direction if the quark has the highest energy, and anti-parallel to the 
anti-quark direction if the anti-quark has the highest energy. 
 The cross-section consists of six terms, each of which may be 
factorized into three contributions: 1) event orientation factor in terms
of $\theta$ and $\chi$; 2) $d^2\sigma_i/dxd{\bar x}$ ($i$=U,..,A) determined
by QCD; and 3) $h_f^{(1,2)}$ determined by the fermion electroweak couplings
and  beam polarization. While the first four terms are P-even, 
the last two terms are P-odd, and are sensitive to any parity-violating
interactions at the $Z^0{\rm q}{\bar {\rm q}}$ and gq${\bar {\rm q}}$ 
vertices. 
In addition to these six terms, the most general differential cross section 
can have three more terms that are odd 
under time-reversal~\cite{Hagiwara}. Being T-odd, however, these terms 
vanish at tree level in a theory that respects CPT invariance.

Recently Burrows and Osland have proposed new QCD tests in terms of the 
event orientation angles~\cite{Phil}.
Integrating over scaled momenta and $\chi$, the polar angle 
distribution of the thrust axis can be expressed as 
\begin{equation}
\sigma(\cos\theta)\equiv {{d\sigma} \over {d\cos\theta}} \propto 
(1-P_{e^-}\cdot A_e)(1+\alpha
\cos^2\theta)+2 A_P (P_{e^-}-A_e) \cos\theta, 
\end{equation}
where  $A_P$ is the parity violation parameter:
\begin{equation}
\label{AP}
A_P = {{\hat{\sigma}_P}\over{\hat{\sigma}_U+\hat{\sigma}_L}} A_f.
\end{equation}
Here $\alpha =
{{\hat{\sigma}_U-2\hat{\sigma}_L}\over{\hat{\sigma}_U+2\hat{\sigma}_L}}$,
 $\hat{\sigma}_i = \int{{d\sigma_i} \over {dxd{\bar x}}}dxd{\bar x}$,   
and $P_{e^-}$ is the signed electron beam polarization, and $A_e ~(A_f)$
is the electroweak coupling of the $Z^0$ to the initial 
(final) state, given by $A_i = 2v_ia_i/(v_i^2+a_i^2)$   in terms of
the vector $v_i$ and axial-vector $a_i$  couplings.
By manipulating the polarization sign of the electron beam
the left-right-forward-backward asymmetry, $\tilde{A}_{FB}$, is directly 
sensitive to 
the asymmetry parameter $A_P$,
$$\tilde{A}_{FB}(|\cos\theta|) \equiv {{\sigma_L(|\cos\theta|)-\sigma_L(-|\cos\theta|)
+ \sigma_R(-|\cos\theta|)-\sigma_R(|\cos\theta|) } \over
{\sigma_L(|\cos\theta|)+\sigma_L(-|\cos\theta|)
+\sigma_R(-|\cos\theta|)+\sigma_R(|\cos\theta|)}}$$  
\begin{equation}
 = |P_e|A_P {{2|\cos\theta|} \over {1+\cos^2\theta}}.
\end{equation} 
Similarly, by integrating over $\cos\theta$, $x$ and $\bar x$, 
the azimuthal-angle distribution can be expressed as
\begin{equation}
2\pi{{d\sigma}\over {d\chi}} \propto (1-P_{e^-}\cdot A_e)(1+\beta
\cos 2\chi) - {3\pi\over {2\sqrt{2}}} A_P'  (P_{e^-}-A_e) \cos\chi, 
\end{equation}
where $A_P'$ is the parity violation parameter:
\begin{equation}
A_P' = {{\hat{\sigma}_A}\over{\hat{\sigma}_U+\hat{\sigma}_L}} A_f,
\end{equation}
with $\beta = {{\hat{\sigma}_T}\over{\hat{\sigma}_U+\hat{\sigma}_L}}$.
Given the value of the electroweak parameter $A_f$, 
measurement of the angular asymmetry parameters $A_P$ and $A_P'$ in 
\z0$\rightarrow$\qqg events allows one to test the 
QCD prediction for $\hat{\sigma}_P/(\hat{\sigma}_U+\hat{\sigma}_L)$ and 
 $\hat{\sigma}_A/(\hat{\sigma}_U+\hat{\sigma}_L)$. Furthermore, the ratio
$A_P/A_P'$ is independent of $A_f$ and is proportional to 
$\hat{\sigma}_A/\hat{\sigma}_P$.
  
The differential cross-section can also be expressed in terms of the polar 
angle $\omega$ of the vector ${\bar n}$ normal to the event plane w.r.t. 
the electron beam direction, where $\cos\omega$=$\sin\theta\sin\chi$:   
\begin{equation}
{{d\sigma} \over {d\cos\omega}} \propto (1-P_{e^-}\cdot A_e)(1+\gamma 
\cos^2\omega)+{16 \over 9} A_T (P_{e^-}-A_e) \cos\omega, 
\end{equation}
with $\gamma = -{1\over3}{{\hat{\sigma}_U-2\hat{\sigma}_L+6\hat{\sigma}_T}\over
{\hat{\sigma}_U+2/3\hat{\sigma}_L+2/3\hat{\sigma}_T}}$.
At first  order in perturbative QCD, $\hat{\sigma}_L$ = 2$\hat{\sigma}_T$,
 yielding $\gamma = -{1\over3}$, and $A_T$ = 0. 
The second term is one of the three T-odd terms mentioned above,
and appears as a forward-backward asymmetry of the
event-plane normal relative to the \z0 polarization axis.
The left-right-forward-backward asymmetry in $\cos\omega$ can also be defined
by a similar double asymmetry as Eq.~ 4, and is directly 
proportional to the T-odd parameter $A_T$.
The vector normal to the event plane can be defined in two ways:
1) the three jets are ordered according to their energies, and the two highest
energy jet momenta are used to define ${\bar n} = \vec{p}_1 \times \vec{p}_2$; 
and 2)  the quark and anti-quark
momenta are used to define ${\bar n} = \vec{p}_q \times \vec{p}_{\bar q}$. 
The asymmetry term
is CP-even in the first definition, and CP-odd in the second.
The first definition does not require jet flavor identification, and
we have studied the asymmetry for inclusive hadronic \z0 decays 
~\cite{t-odd}. 
The second definition requires tagging both  quark- and antiquark-jets.
In both cases, in the Standard Model the asymmetry vanishes identically at 
tree level, but higher-order processes  yield non-zero contributions for 
\ep$\rightarrow$\bbg. However,
due to various cancellations, these contributions are found to be very small
at the \z0 resonance and yield values of the asymmetry parameter
$|A_T| < 10^{-5}$~\cite{Brandenburg}.
Measurement of the asymmetry in $\cos\omega$ is hence potentially sensitive to
physics processes beyond the Standard Model~\cite{Murayama}.   
 
\section{Event and Track Selection }

\noindent 
The SLAC Linear Collider (SLC) collides
longitudinally polarized electrons with unpolarized positrons
at a center-of-mass energy of 91.2 GeV.
The electron polarization direction is randomly reversed  pulse-by-pulse, 
reducing systematic effects on polarization-dependent asymmetries.
The magnitude of the average electron-beam
polarization was 0.63 in 1993, 0.77 in 1994-1996, and 0.73 in 1997-1998.
  
 The measurement was performed with the SLC Large Detector (SLD) using
approximately 550,000 \z0 decays collected between
 1993 and 1998. A general description of the SLD can
be found elsewhere~\cite{sld}. Charged particle tracking and momentum 
analysis is provided by the central drift chamber (CDC)~\cite{cdc} 
and the CCD-based vertex detector (VXD)~\cite{vxd} 
in a uniform axial magnetic field of 0.6 T.
About 70\% of the data were taken with a new vertex detector (VXD3) installed
in 1996, and the rest with the previous detector, VXD2.
Particle energies are measured in the liquid argon calorimeter
(LAC)~\cite{LAC} and in the warm iron calorimeter~\cite{WIC}.

In the present analysis the hadronic event selection, three-jet
reconstruction, and b-tagging
were based on charged tracks. A set of cuts was applied 
to the data to select well-measured tracks and events well contained
within the detector acceptance~\cite{had}. 
Events were required to have (i) at least 7 charged tracks; (ii) a visible 
charged energy of at least 20 GeV; and (iii) a thrust axis
~\cite{thrust} polar angle
satisfying $|\cos\theta_T|<$0.71, which was reconstructed using the LAC.
Charged tracks reconstructed in the CDC were linked with pixel clusters
 in the VXD by extrapolating each track and selecting the best set of 
associated clusters.
The average efficiency of reconstruction in the CDC
and linking to the correct set of VXD hits is 95\% (94\%) for the region 
$|\cos\theta|<$ 0.85 (0.74)~\cite{vxd3}.
The momentum resolution of the combined CDC and VXD systems is
$(\delta p _{\perp} / p _{\perp})^2 = (.01)^2 + (.0026p_{\perp})^2$,
where $p_{\perp}$ is the transverse  momentum in GeV/c w.r.t. the beamline.

The centroid of the micron-size SLC Interaction Point (IP) in the $r\phi$
plane is reconstructed with a measured precision of $\sigma_{r\phi}^{IP}$ 
$\sim$ 5$\mu$m (7$\mu$m)
using tracks in sets of $\approx$30 sequential hadronic
\z0 decays. The $z$ position of the IP is determined on an event-by-event
basis with a precision of $\sigma_z^{IP}\sim$32$\mu$m (52$\mu$m) 
using the median $z$ 
position of tracks at their point-of-closest approach to the IP in the
$r\phi$ plane. The track impact parameter resolution at high
momentum is 11$\mu$m (11$\mu$m) in the plane perpendicular to the beam axis 
($r\phi$ plane) and 
23$\mu$m (38$\mu$m) in the plane containing the beam axis ($rz$ plane).

A set of ``quality'' tracks for use in heavy quark tagging 
 was selected. Tracks measured in the CDC were required to
have $\ge$40 hits, with the first hit at a radius $r <$ 39 cm, 
a transverse momentum $p_{\perp} > $ 0.4 GeV/c, a good fit quality
($\chi^2/N_{DOF} <$ 5), and to extrapolate to the IP within 1 cm in
$r\phi$ and 1.5 cm in $z$. Tracks were required to have at least one
associated VXD hit, and a combined CDC-VXD fit with $\chi^2/N_{DOF} <$ 5.
Tracks with an $r\phi$ impact parameter $\delta >$ 3 mm or with
an impact parameter error $\sigma_\delta >$ 250 $\mu$m were removed.
Tracks from identified $\gamma$ conversions and $K^0$ or $\Lambda^0$ decays
were also removed. 
 
%------------------------------------------------------
% Analysis
%------------------------------------------------------
\section{\bbg Analysis}
 
 Three-jet
events were selected and the three momentum vectors of the jets were
 reconstructed.
Although the parton momenta are not directly measurable, at $\sqrt{s}$
$\approx$ 91 GeV the partons usually
appear as well-collimated jets of hadrons. Jets were
reconstructed using the ``Durham'' jet algorithm \cite{durham}.
Planar three-jet events were
selected by requiring exactly three reconstructed jets to be found with a
jet-resolution parameter value of
$y_c$=0.005, the sum of the angles
between the three jets to be greater than 358$^\circ$,
and that each jet contain at least two charged tracks.
A total of 75,000 events satisfied these
criteria.
 
Such jet algorithms accurately reconstruct
the parton directions
but measure the parton energies poorly \cite{threejet}.
Therefore, the jet energies were
calculated by using the measured jet directions and solving the
three-body kinematics assuming massless jets, and
were then used to label the jets such that $E_1 > E_2 > E_3$.
 The energy of jet 1, for example, is given by
 
\begin{equation}
E_1 = \sqrt{s} {{\sin\theta_{23}} \over
 {\sin\theta_{12}+\sin\theta_{23}+\sin\theta_{31}}},
\label{energy}
\end{equation}
 
\noindent
where $\theta_{kl}$ is the angle between jets $k$ and $l$.

To select \bbg events the long lifetime and large invariant mass of B-hadrons 
was exploited.
A topological algorithm ~\cite{zvtop} was applied to the set of
quality tracks in each  jet to search for a secondary decay vertex. 
Vertices were required to be separated from the
IP by at least 1 mm and to contain at least two tracks.
Monte Carlo studies show that
the probability for reconstructing at least one such vertex was
$\sim$ 91\% (77\%) in \bbg events, $\sim$ 45\% (26\%) in \ccg events, and
$\sim$~2\% (2\%) in light quark events. Once a vertex was found, additional
 tracks consistent with coming from the vertex were attached 
 in an attempt to reconstruct the invariant mass of a B-hadron.
A vertex axis was formed by a straight line joining the IP and the
 vertex, which was located at a distance $D$ from the IP.
For each quality track
the distance of closest approach, $T$, and the distance from the IP
along the vertex axis to the point of closest approach, $L$, were calculated.
Tracks with $T <$ 1 mm, and $L/D >$ 0.25 were attached 
to the secondary vertex, and the vertex invariant mass, $M_{ch}$, was 
calculated assuming each track was a charged pion. Due to neutral decay
products the total momentum vector of the tracks 
and the vertex axis were typically
acollinear. To account for the missing neutral particles, 
an additional component of  transverse 
momentum $P_t$, defined by the projection of the total momentum vector 
perpendicular to the
vertex axis, was added to yield $M=\sqrt{M_{ch}^2+P_t^2}+|P_t|$~\cite{miss}. 
Figures 1(a), (b), and (c) show the distributions of this $P_t$-corrected
  vertex mass for jet 1, 2, and  3, respectively. 
An event was selected as \bbg if at least one jet contained a vertex with 
$M >$ 1.5 GeV/c$^2$. 
A total of 14,658 events satisfied 
this requirement and were subjected to further analysis.
Monte Carlo studies show that this selection is 84\% (69\%) efficient for 
identifying a sample of \bbg events with 84\% (87\%) purity, and 
containing 14\% (11\%)
 $c{\bar {\rm c}}$g and 2\% (2\%) light-flavor backgrounds.     

The identification of each jet was based on  the momentum-weighted 
jet charge and $r\phi$ impact parameter techniques. 
The momentum-weighted  charge was calculated for each jet:
\begin{equation}
 Q_j = \sum q_i |\vec{p}_i\cdot \vec{\hat{t}}|^\kappa,
\end{equation}
where $\kappa$=0.5, $\vec{\hat{t}}$ is the unit vector along the event
thrust axis, and $q_i$ and $\vec{p}_i$ are the
charge and momentum of the $i^{th}$ track associated with jet $j$.
We then examined  the difference in the momentum-weighted jet charge, 
$Q_{diff}$ = $Q_1 - Q_2 - Q_3$. If this quantity was  
 negative (positive), jet 1 was tagged as the b-jet
(${\bar {\rm b}}$-jet).
The jet flavor was tagged 
by counting the number of ``significant'' tracks with normalized  impact
parameter w.r.t the IP $d/\sigma_d >$ 3.  Figures 2(a), (b), and (c) show 
the distributions of the number of significant tracks found in
jets 1, 2, and 3, respectively, in the b-tagged events.
Jet 1 was chosen as the gluon-jet only if jet 1 had no significant track and
both jet 2 and 3 had at least one significant track. Jet 2 was chosen
as the gluon-jet if jet 2 had no significant track and jet 3 had at least
one significant track. Otherwise, jet 3 was chosen as the gluon-jet.

\section{Monte Carlo Simulation}
 
A Monte Carlo simulation of hadronic \z0
decays combined with a simulation of the detector
response was used to 
study the quality of the jet reconstruction, the b-tagging efficiency 
and purity, and the efficiency of the jet flavor identification.
The JETSET 7.4~\cite{JETSET} event generator was used, with parameter 
values tuned to hadronic \ep annihilation data~\cite{PNB}, combined
with a simulation of $B$ hadron decays tuned~\cite{PUB7117} to $\Upsilon(4S)$
data and a detector simulation based on GEANT 3.21~\cite{GEANT}.  
For those events satisfying the three-jet criteria,
exactly three jets were
reconstructed at the parton level by applying the jet algorithm to
the parton momenta. The
three parton-level jets were associated with the three detector-level
jets by choosing the
combination that minimized the sum of the angular differences between the
 corresponding jets, and the energies and charges of the matching
jets were compared.

For the T-odd asymmetry analyses the vector normal to the jet plane
is measured in two ways: 1) using the two highest energy jets, and 
2) using identified b- and \bbar-jets. 
In the first method, where the jets are labeled according to their energy,
six detector-jet energy orderings are possible for a given parton-jet
energy ordering. For the three cases where the energy ordering of
any two jets does not agree between parton and detector levels,
the direction of the jet-plane normal vector is opposite between 
the parton level and detector level and $\cos\omega$  will be
measured with the wrong sign. The average probability of measuring 
$\cos\omega$ with the correct sign in this analysis is estimated from 
the simulation to be 76\% (76\%). 
In the second method, where both b- and \bbar-jets
are identified, the gluon-jet must be tagged correctly, and
furthermore, the charge assignment of the  b- and \bbar-jets must be
correct. The average probability of 
identifying the gluon-jet correctly is 91\% (88\%), 
and combined with the correct-charge assignment
probability  determined by the self-calibration technique described 
in the next section,
the average probability of measuring $\cos\omega$
with the right sign is 64\% (63\%). 

\section{Angular Asymmetries}

Figures 3(a) and (b) show the observed $\cos\theta$ distributions of the 
signed-thrust axis for event samples collected with 
 left- and right-handed electron beam, respectively.
The histograms show the backgrounds estimated using the simulation.
The thrust axis is signed so that it points towards the hemisphere
containing the b-tagged jet.
The $\cos\theta$ distribution may be described by

\begin{eqnarray}
\lefteqn{{{d\sigma} \over {d\cos\theta}} =(1-P_{e^-}\cdot A_e)(1 + \alpha 
\cos^2\theta)~ + \hspace{1in} }   \nonumber\\
& & \qquad 2\,(P_{e^-}-A_e)\,\cos\theta~[ A_{P,b}\,f_b\,(2\,p^{correct,b}-1) + 
\nonumber\\
& &  \qquad A_{P,c}\,f_c\,(2\,p^{correct,c}-1) + A_{P,uds}\,(1-f_b-f_c)\,
(2\,p^{correct,uds}-1)~], 
\label{diffb}
\end{eqnarray}
where $f_b$, $f_c$, $f_{uds}$ are the fractions of \bbg, \ccg, and
light quark events in the sample, determined from the Monte Carlo 
simulation, and
 $p^{correct,b}$, $p^{correct,c}$, $p^{correct,uds}$ are the
probabilities to tag the parton charge correctly for \bbg, \ccg, and the light
quark events, respectively. The correct-charge probability is calculated 
 as a function of the measured jet charge difference $|Q_{diff}|$ as
$p^{correct} = 1 / (1+e^{-\alpha |Q_{diff}|})$.
The quantity $\alpha$ is a parametrization of how well the 
momentum-weighted charge technique signs the thrust axis direction.
While $\alpha_c$ and $\alpha_{uds}$ for \ccg and light-quark backgrounds
were calculated from the simulation, $\alpha_b$ for \bbg events was
determined from data using a self-calibration technique~\cite{junk}. 
Using the measured widths, $\sigma_{diff}$ and $\sigma_{sum}$, of the
$Q_{diff}~ (= Q_1 - Q_2 - Q_3)$ and $Q_{sum} ~(= Q_1 + Q_2 + Q_3)$
 distributions:
\begin{equation}
\alpha_b = \frac{2\sqrt{\sigma^2_{diff}-(1+\lambda)^2\sigma^2_{sum}}}
               {(1+\lambda)^2\sigma^2_{sum}},
\end{equation}
where  the hemisphere correlation $\lambda$ = 0.027 was calculated
 from the simulation. This  yielded $\alpha_b$ = 0.218$\pm$0.021 
(0.255$\pm$0.032) averaged over 
$\cos\theta$.
On average the correct-charge assignment probability for \bbg events is 
68\% (67\%).
The  asymmetry parameters $A_{P,c}$ and $A_{P,uds}$ 
(Eq.~\ref{AP})
for charm and light-quark backgrounds were calculated from the simulation 
based on the Standard Model. 
A  maximum-likelihood
fit of Eq.~\ref{diffb} is performed to extract $A_{P,b}$. We found
\begin{equation}
A_{P,b} = 0.847 \pm 0.049, \quad (PRELIMINARY)
\label{Ab}
\end{equation} 

\noindent
where the error is statistical only. Assuming the Standard Model expectation 
of $A_b$ = 0.94 for $\sin^2\theta_w$=0.23, the measured value of $A_{P,b}$ 
yields
$${{\hat{\sigma}_P} \over {\hat{\sigma}_U+\hat{\sigma}_L}} 
= 0.906 \pm 0.052 (stat.).$$
This value is consistent with the ${\cal O}(\alpha_s^2)$ QCD expectation of
 $\hat{\sigma}_P/(\hat{\sigma}_U+\hat{\sigma}_L)$ = 0.93, calculated using 
the JETSET 7.4 event generator~\cite{JETSET}.    

Figures 4(a) and (b) show the $\chi$ distributions for event samples collected
with left- and right-handed electron beam, respectively.
The $\chi$ distribution may be described by

\begin{eqnarray}
\lefteqn{{{d\sigma} \over {d\chi}} =(1-P_{e^-}\cdot A_e)(1 + \beta 
\cos 2\chi)~ - \hspace{1in} }   \nonumber\\
& & \qquad {3\pi \over{2\sqrt{2}}}\,(P_{e^-}-A_e)\,\cos\chi~
[ A_{P,b}'\,f_b\,P_{AP}^b + 
\nonumber\\
& &  \qquad A_{P,c}'\,f_c\,P_{AP}^c + A_{P,uds}'\,(1-f_b-f_c)\,P_{AP}^{uds}~], 
\label{diffc}
\end{eqnarray}

\noindent
where $P_{AP}^b$, $P_{AP}^c$, and $P_{AP}^{uds}$ are the analyzing powers 
for \bbg,
\ccg, and light quark events, respectively, and are function of the probability
to tag the parton charge correctly, $P_{chg}$, and the probability to tag the 
gluon-jet correctly, $P_{glu}$, given by

\begin{equation}
 P_{AP} = P_{chg} P_{glu} -  
(1-P_{chg})P_{glu} - P_{chg} (1-P_{glu}) + (1-P_{chg})(1-P_{glu}). 
\end{equation}
A  maximum-likelihood
fit of Eq.~\ref{diffc} is performed to extract $A_{P,b}'$. We found
\begin{equation}
A_{P,b}' = -0.013 \pm 0.033, \quad (PRELIMINARY)
\label{Ab}
\end{equation} 

\noindent
where the error is statistical only. Assuming the Standard Model expectation 
of $A_b$ = 0.94 for $\sin^2\theta_w$=0.23, the measured value of $A_{P,b}'$ 
yields
$${{\hat{\sigma}_A} \over {\hat{\sigma}_U+\hat{\sigma}_L}} 
= -0.014 \pm 0.035 (stat.).$$
This value is consistent with the ${\cal O}(\alpha_s^2)$ QCD expectation of
 $\hat{\sigma}_A/(\hat{\sigma}_U+\hat{\sigma}_L)$ = $-$0.064, calculated using 
the JETSET 7.4 event generator~\cite{JETSET}.
    
Figures 5(a) and (b) show the left-right-forward-backward asymmetry of the
$\cos\omega$ distribution for the two definitions:
 (a) $\vec{p_1}\times\vec{p_2}$, and (b) $\vec{p_b}\times\vec{p_{\bar b}}$. 
No asymmetry is apparent. The $\cos\omega$ distribution may be described,
assuming no asymmetries in the \ccg and light-quark backgrounds, by
 
\begin{equation}
\label{t-odd-fit}
{{d\sigma} \over {d\cos\omega}} \propto (1-P_{e^-}\cdot A_e)(1 - {1\over3} 
\cos^2\omega)+{16 \over 9}\,(P_{e^-}-A_e)\,A_T\, f_b\,P_{AP}\,\cos\omega, 
\label{diffd}
\end{equation}

\noindent
where $f_b$ is the fraction of \bbg events in the sample, and
 the analyzing power, $P_{AP}$, represents the probability of correctly signing 
the vector normal to the event plane. In the first case $P_{AP}$ 
is given by the probability of
correct energy-ordering, $P_{AP} = (2p^{correct} -1)$, and in the second case 
it is the probability of correct-sign assignment combined with the tagging 
efficiency of the gluon-jet.   
We performed maximum-likelihood fits of Eq.~\ref{t-odd-fit} to the
$\cos\omega$ distributions to extract the parameters 
$A_T^+$, for the CP-even case, and $A_T^-$, for the CP-odd case.

 We found
$$ A_T^+ = -0.012 \pm 0.013, \quad (PRELIMINARY)$$
$$ A_T^- = -0.033 \pm 0.023, \quad (PRELIMINARY)$$
where the error is statistical only. In both cases the T-odd contribution is 
consistent with zero within the statistical error and we calculate limits of
$$-0.038 < A_T^+ < 0.014 \quad @ \quad 95\% \quad C.L., \quad (PRELIMINARY)$$
$$-0.077 < A_T^- < 0.011 \quad @ \quad 95\% \quad C.L..  \quad (PRELIMINARY)$$
The results of these fits are shown in Figures~5(a) and 5(b).

%------------------------------------------------------
% Systematics
%-----------------------------------------------------
\section{Systematic Errors}
 
Table~\ref{sys} summarizes the systematic errors on the forward-backward 
asymmetry analysis of the signed thrust-axis.
The largest systematic error was due to the statistical uncertainty in 
the $\alpha_b$ determination using the self-calibration technique.
This error would decrease with a larger data sample. 
The systematic error in  the inter-hemisphere correlation $\lambda$ was 
due to the limited statistics of the Monte Carlo simulation.
The systematic error in the tag composition was due to the heavy quark 
physics modeling. In \bb events we have considered the uncertainties
on: the branching fraction for \z0$\rightarrow$\bb, the $B$ hadron 
fragmentation function, the rates of production
of $B^\pm$, $B^0$ and $B^0_s$ mesons, and $B$ baryons, the lifetimes of $B$
mesons and baryons, and the average $B$ hadron decay charge multiplicity.
In \cc events we have considered the uncertainties on: the branching
fraction for \z0$\rightarrow$\cc, the charmed hadron fragmentation
function, the rates of production of $D^0$, $D^+$ and $D_s$ mesons, and
charmed baryons, and the charged multiplicity of charmed hadron decays.  
We have also considered the rate of production of $s\bar{s}$ in the jet
fragmentation process, and the production of secondary
\bb and \cc from gluon splitting.
The systematic error in the detector modeling results from
discrepancies between data and Monte Carlo in 
tracking efficiency and resolution. 
The systematic errors on $A_P'$ and $A_T$ are negligibly small as the
uncertainty diminishes with the asymmetry itself. 

%------------------------------------------------------
% Conclusions
%------------------------------------------------------
%==============
% Conclusions
%==============
\section{Conclusions}
 
In conclusion, we have made the first angular asymmetry measurements in
polarized \z0 decays to \bbg. 
From the forward-backward polar angle asymmetry of the signed-thrust axis
we have measured the parity violation parameter 
$A_P$ = 0.847 $\pm$ 0.049 (stat.) $\pm$ 0.060 (syst.).  
From the azimuthal angle asymmetry we have measured the second parity
violation parameter
$A_P'$ = $-$0.013 $\pm$ 0.033 (stat.) $\pm$ 0.002 (syst.). 
Assuming the Standard Model expectation of $A_b$ = 0.94, 
the QCD factors for \bbg events are measured to be
 $\hat{\sigma}_P/(\hat{\sigma}_U+\hat{\sigma}_L)$ =  0.906 $\pm$0.052 (stat.)
$\pm$ 0.064 (syst.), and  
 $\hat{\sigma}_A/(\hat{\sigma}_U+\hat{\sigma}_L)$ =  $-$0.014 $\pm$0.035 (stat.)
$\pm$ 0.002 (syst.),
which are consistent with the ${\cal O}(\alpha_s^2)$ QCD
expectations. 
We find the T-odd asymmetry to be consistent with zero, and  we set
95\% C.L. limits on the asymmetry parameter 
$-$0.038 $< A_T^+ <$ 0.014 for the CP-even case and 
$-$0.077 $< A_T^- <$ 0.011 for the CP-odd case. All results are preliminary.

\section*{Acknowledgements}
We thank the personnel of the SLAC accelerator department and the
technical
staffs of our collaborating institutions for their outstanding efforts
on our behalf.

\vskip .5truecm

\vbox{\footnotesize\renewcommand{\baselinestretch}{1}\noindent
$^*$Work supported by Department of Energy
  contracts:
  DE-FG02-91ER40676 (BU),
  DE-FG03-91ER40618 (UCSB),
  DE-FG03-92ER40689 (UCSC),
  DE-FG03-93ER40788 (CSU),
  DE-FG02-91ER40672 (Colorado),
  DE-FG02-91ER40677 (Illinois),
  DE-AC03-76SF00098 (LBL),
  DE-FG02-92ER40715 (Massachusetts),
  DE-FC02-94ER40818 (MIT),
  DE-FG03-96ER40969 (Oregon),
  DE-AC03-76SF00515 (SLAC),
  DE-FG05-91ER40627 (Tennessee),
  DE-FG02-95ER40896 (Wisconsin),
  DE-FG02-92ER40704 (Yale);
  National Science Foundation grants:
  PHY-91-13428 (UCSC),
  PHY-89-21320 (Columbia),
  PHY-92-04239 (Cincinnati),
  PHY-95-10439 (Rutgers),
  PHY-88-19316 (Vanderbilt),
  PHY-92-03212 (Washington);
  The UK Particle Physics and Astronomy Research Council
  (Brunel, Oxford and RAL);
  The Istituto Nazionale di Fisica Nucleare of Italy
  (Bologna, Ferrara, Frascati, Pisa, Padova, Perugia);
  The Japan-US Cooperative Research Project on High Energy Physics
  (Nagoya, Tohoku);
  The Korea Research Foundation (Soongsil, 1997).}

\vskip 1truecm
  
\section*{$^{**}$List of Authors}
%%%%%%%%% sld author list starts here
%
% author list for inclusion in LaTeX documents
% using \author{} and \address{} commands
%
% Institution number definitions:
%
\begin{center}
\def\iADEL{$^{(1)}$}
\def\iAOMORI{$^{(2)}$}
\def\iBOLO{$^{(3)}$}
\def\iBRI{$^{(4)}$}
\def\iBRUN{$^{(5)}$}
\def\iBU{$^{(6)}$}
\def\iCINC{$^{(7)}$}
\def\iCOLO{$^{(8)}$}
\def\iCOLU{$^{(9)}$}
\def\iCSU{$^{(10)}$}
\def\iFERR{$^{(11)}$}
\def\iFRAS{$^{(12)}$}
\def\iILLI{$^{(13)}$}
\def\iJHU{$^{(14)}$}
\def\iLBL{$^{(15)}$}
\def\iLTU{$^{(16)}$}
\def\iMASS{$^{(17)}$}
\def\iMISSI{$^{(18)}$}
\def\iMIT{$^{(19)}$}
\def\iMOSCOW{$^{(20)}$}
\def\iNAGO{$^{(21)}$}
\def\iOREG{$^{(22)}$}
\def\iOXF{$^{(23)}$}
\def\iPADO{$^{(24)}$}
\def\iPERU{$^{(25)}$}
\def\iPISA{$^{(26)}$}
\def\iRAL{$^{(27)}$}
\def\iRUTG{$^{(28)}$}
\def\iSLAC{$^{(29)}$}
\def\iSOGA{$^{(30)}$}
\def\iSOONG{$^{(31)}$}
\def\iTENN{$^{(32)}$}
\def\iTOHO{$^{(33)}$}
\def\iUCSB{$^{(34)}$}
\def\iUCSC{$^{(35)}$}
\def\iUVIC{$^{(36)}$}
\def\iVAND{$^{(37)}$}
\def\iWASH{$^{(38)}$}
\def\iWISC{$^{(39)}$}
\def\iYALE{$^{(40)}$}

  \baselineskip=.75\baselineskip  
\mbox{Kenji  Abe\unskip,\iNAGO}
\mbox{Koya Abe\unskip,\iTOHO}
\mbox{T. Abe\unskip,\iSLAC}
\mbox{I.Adam\unskip,\iSLAC}
\mbox{T.  Akagi\unskip,\iSLAC}
\mbox{N. J. Allen\unskip,\iBRUN}
\mbox{W.W. Ash\unskip,\iSLAC}
\mbox{D. Aston\unskip,\iSLAC}
\mbox{K.G. Baird\unskip,\iMASS}
\mbox{C. Baltay\unskip,\iYALE}
\mbox{H.R. Band\unskip,\iWISC}
\mbox{M.B. Barakat\unskip,\iLTU}
\mbox{O. Bardon\unskip,\iMIT}
\mbox{T.L. Barklow\unskip,\iSLAC}
\mbox{G. L. Bashindzhagyan\unskip,\iMOSCOW}
\mbox{J.M. Bauer\unskip,\iMISSI}
\mbox{G. Bellodi\unskip,\iOXF}
\mbox{R. Ben-David\unskip,\iYALE}
\mbox{A.C. Benvenuti\unskip,\iBOLO}
\mbox{G.M. Bilei\unskip,\iPERU}
\mbox{D. Bisello\unskip,\iPADO}
\mbox{G. Blaylock\unskip,\iMASS}
\mbox{J.R. Bogart\unskip,\iSLAC}
\mbox{G.R. Bower\unskip,\iSLAC}
\mbox{J. E. Brau\unskip,\iOREG}
\mbox{M. Breidenbach\unskip,\iSLAC}
\mbox{W.M. Bugg\unskip,\iTENN}
\mbox{D. Burke\unskip,\iSLAC}
\mbox{T.H. Burnett\unskip,\iWASH}
\mbox{P.N. Burrows\unskip,\iOXF}
\mbox{A. Calcaterra\unskip,\iFRAS}
\mbox{D. Calloway\unskip,\iSLAC}
\mbox{B. Camanzi\unskip,\iFERR}
\mbox{M. Carpinelli\unskip,\iPISA}
\mbox{R. Cassell\unskip,\iSLAC}
\mbox{R. Castaldi\unskip,\iPISA}
\mbox{A. Castro\unskip,\iPADO}
\mbox{M. Cavalli-Sforza\unskip,\iUCSC}
\mbox{A. Chou\unskip,\iSLAC}
\mbox{E. Church\unskip,\iWASH}
\mbox{H.O. Cohn\unskip,\iTENN}
\mbox{J.A. Coller\unskip,\iBU}
\mbox{M.R. Convery\unskip,\iSLAC}
\mbox{V. Cook\unskip,\iWASH}
\mbox{R. Cotton\unskip,\iBRUN}
\mbox{R.F. Cowan\unskip,\iMIT}
\mbox{D.G. Coyne\unskip,\iUCSC}
\mbox{G. Crawford\unskip,\iSLAC}
\mbox{C.J.S. Damerell\unskip,\iRAL}
\mbox{M. N. Danielson\unskip,\iCOLO}
\mbox{M. Daoudi\unskip,\iSLAC}
\mbox{N. de Groot\unskip,\iBRI}
\mbox{R. Dell'Orso\unskip,\iPERU}
\mbox{P.J. Dervan\unskip,\iBRUN}
\mbox{R. de Sangro\unskip,\iFRAS}
\mbox{M. Dima\unskip,\iCSU}
\mbox{A. D'Oliveira\unskip,\iCINC}
\mbox{D.N. Dong\unskip,\iMIT}
\mbox{M. Doser\unskip,\iSLAC}
\mbox{R. Dubois\unskip,\iSLAC}
\mbox{B.I. Eisenstein\unskip,\iILLI}
\mbox{V. Eschenburg\unskip,\iMISSI}
\mbox{E. Etzion\unskip,\iWISC}
\mbox{S. Fahey\unskip,\iCOLO}
\mbox{D. Falciai\unskip,\iFRAS}
\mbox{C. Fan\unskip,\iCOLO}
\mbox{J.P. Fernandez\unskip,\iUCSC}
\mbox{M.J. Fero\unskip,\iMIT}
\mbox{K.Flood\unskip,\iMASS}
\mbox{R. Frey\unskip,\iOREG}
\mbox{J. Gifford\unskip,\iUVIC}
\mbox{T. Gillman\unskip,\iRAL}
\mbox{G. Gladding\unskip,\iILLI}
\mbox{S. Gonzalez\unskip,\iMIT}
\mbox{E. R. Goodman\unskip,\iCOLO}
\mbox{E.L. Hart\unskip,\iTENN}
\mbox{J.L. Harton\unskip,\iCSU}
\mbox{A. Hasan\unskip,\iBRUN}
\mbox{K. Hasuko\unskip,\iTOHO}
\mbox{S. J. Hedges\unskip,\iBU}
\mbox{S.S. Hertzbach\unskip,\iMASS}
\mbox{M.D. Hildreth\unskip,\iSLAC}
\mbox{J. Huber\unskip,\iOREG}
\mbox{M.E. Huffer\unskip,\iSLAC}
\mbox{E.W. Hughes\unskip,\iSLAC}
\mbox{X.Huynh\unskip,\iSLAC}
\mbox{H. Hwang\unskip,\iOREG}
\mbox{M. Iwasaki\unskip,\iOREG}
\mbox{D. J. Jackson\unskip,\iRAL}
\mbox{P. Jacques\unskip,\iRUTG}
\mbox{J.A. Jaros\unskip,\iSLAC}
\mbox{Z.Y. Jiang\unskip,\iSLAC}
\mbox{A.S. Johnson\unskip,\iSLAC}
\mbox{J.R. Johnson\unskip,\iWISC}
\mbox{R.A. Johnson\unskip,\iCINC}
\mbox{T. Junk\unskip,\iSLAC}
\mbox{R. Kajikawa\unskip,\iNAGO}
\mbox{M. Kalelkar\unskip,\iRUTG}
\mbox{Y. Kamyshkov\unskip,\iTENN}
\mbox{H.J. Kang\unskip,\iRUTG}
\mbox{I. Karliner\unskip,\iILLI}
\mbox{H. Kawahara\unskip,\iSLAC}
\mbox{Y. D. Kim\unskip,\iSOGA}
\mbox{M.E. King\unskip,\iSLAC}
\mbox{R. King\unskip,\iSLAC}
\mbox{R.R. Kofler\unskip,\iMASS}
\mbox{N.M. Krishna\unskip,\iCOLO}
\mbox{R.S. Kroeger\unskip,\iMISSI}
\mbox{M. Langston\unskip,\iOREG}
\mbox{A. Lath\unskip,\iMIT}
\mbox{D.W.G. Leith\unskip,\iSLAC}
\mbox{V. Lia\unskip,\iMIT}
\mbox{C.Lin\unskip,\iMASS}
\mbox{M.X. Liu\unskip,\iYALE}
\mbox{X. Liu\unskip,\iUCSC}
\mbox{M. Loreti\unskip,\iPADO}
\mbox{A. Lu\unskip,\iUCSB}
\mbox{H.L. Lynch\unskip,\iSLAC}
\mbox{J. Ma\unskip,\iWASH}
\mbox{G. Mancinelli\unskip,\iRUTG}
\mbox{S. Manly\unskip,\iYALE}
\mbox{G. Mantovani\unskip,\iPERU}
\mbox{T.W. Markiewicz\unskip,\iSLAC}
\mbox{T. Maruyama\unskip,\iSLAC}
\mbox{H. Masuda\unskip,\iSLAC}
\mbox{E. Mazzucato\unskip,\iFERR}
\mbox{A.K. McKemey\unskip,\iBRUN}
\mbox{B.T. Meadows\unskip,\iCINC}
\mbox{G. Menegatti\unskip,\iFERR}
\mbox{R. Messner\unskip,\iSLAC}
\mbox{P.M. Mockett\unskip,\iWASH}
\mbox{K.C. Moffeit\unskip,\iSLAC}
\mbox{T.B. Moore\unskip,\iYALE}
\mbox{M.Morii\unskip,\iSLAC}
\mbox{D. Muller\unskip,\iSLAC}
\mbox{V.Murzin\unskip,\iMOSCOW}
\mbox{T. Nagamine\unskip,\iTOHO}
\mbox{S. Narita\unskip,\iTOHO}
\mbox{U. Nauenberg\unskip,\iCOLO}
\mbox{H. Neal\unskip,\iSLAC}
\mbox{M. Nussbaum\unskip,\iCINC}
\mbox{N.Oishi\unskip,\iNAGO}
\mbox{D. Onoprienko\unskip,\iTENN}
\mbox{L.S. Osborne\unskip,\iMIT}
\mbox{R.S. Panvini\unskip,\iVAND}
\mbox{C. H. Park\unskip,\iSOONG}
\mbox{T.J. Pavel\unskip,\iSLAC}
\mbox{I. Peruzzi\unskip,\iFRAS}
\mbox{M. Piccolo\unskip,\iFRAS}
\mbox{L. Piemontese\unskip,\iFERR}
\mbox{K.T. Pitts\unskip,\iOREG}
\mbox{R.J. Plano\unskip,\iRUTG}
\mbox{R. Prepost\unskip,\iWISC}
\mbox{C.Y. Prescott\unskip,\iSLAC}
\mbox{G.D. Punkar\unskip,\iSLAC}
\mbox{J. Quigley\unskip,\iMIT}
\mbox{B.N. Ratcliff\unskip,\iSLAC}
\mbox{T.W. Reeves\unskip,\iVAND}
\mbox{J. Reidy\unskip,\iMISSI}
\mbox{P.L. Reinertsen\unskip,\iUCSC}
\mbox{P.E. Rensing\unskip,\iSLAC}
\mbox{L.S. Rochester\unskip,\iSLAC}
\mbox{P.C. Rowson\unskip,\iCOLU}
\mbox{J.J. Russell\unskip,\iSLAC}
\mbox{O.H. Saxton\unskip,\iSLAC}
\mbox{T. Schalk\unskip,\iUCSC}
\mbox{R.H. Schindler\unskip,\iSLAC}
\mbox{B.A. Schumm\unskip,\iUCSC}
\mbox{J. Schwiening\unskip,\iSLAC}
\mbox{S. Sen\unskip,\iYALE}
\mbox{V.V. Serbo\unskip,\iSLAC}
\mbox{M.H. Shaevitz\unskip,\iCOLU}
\mbox{J.T. Shank\unskip,\iBU}
\mbox{G. Shapiro\unskip,\iLBL}
\mbox{D.J. Sherden\unskip,\iSLAC}
\mbox{K. D. Shmakov\unskip,\iTENN}
\mbox{C. Simopoulos\unskip,\iSLAC}
\mbox{N.B. Sinev\unskip,\iOREG}
\mbox{S.R. Smith\unskip,\iSLAC}
\mbox{M. B. Smy\unskip,\iCSU}
\mbox{J.A. Snyder\unskip,\iYALE}
\mbox{H. Staengle\unskip,\iCSU}
\mbox{A. Stahl\unskip,\iSLAC}
\mbox{P. Stamer\unskip,\iRUTG}
\mbox{H. Steiner\unskip,\iLBL}
\mbox{R. Steiner\unskip,\iADEL}
\mbox{M.G. Strauss\unskip,\iMASS}
\mbox{D. Su\unskip,\iSLAC}
\mbox{F. Suekane\unskip,\iTOHO}
\mbox{A. Sugiyama\unskip,\iNAGO}
\mbox{S. Suzuki\unskip,\iNAGO}
\mbox{M. Swartz\unskip,\iJHU}
\mbox{A. Szumilo\unskip,\iWASH}
\mbox{T. Takahashi\unskip,\iSLAC}
\mbox{F.E. Taylor\unskip,\iMIT}
\mbox{J. Thom\unskip,\iSLAC}
\mbox{E. Torrence\unskip,\iMIT}
\mbox{N. K. Toumbas\unskip,\iSLAC}
\mbox{T. Usher\unskip,\iSLAC}
\mbox{C. Vannini\unskip,\iPISA}
\mbox{J. Va'vra\unskip,\iSLAC}
\mbox{E. Vella\unskip,\iSLAC}
\mbox{J.P. Venuti\unskip,\iVAND}
\mbox{R. Verdier\unskip,\iMIT}
\mbox{P.G. Verdini\unskip,\iPISA}
\mbox{D. L. Wagner\unskip,\iCOLO}
\mbox{S.R. Wagner\unskip,\iSLAC}
\mbox{A.P. Waite\unskip,\iSLAC}
\mbox{S. Walston\unskip,\iOREG}
\mbox{J.Wang\unskip,\iSLAC}
\mbox{S.J. Watts\unskip,\iBRUN}
\mbox{A.W. Weidemann\unskip,\iTENN}
\mbox{E. R. Weiss\unskip,\iWASH}
\mbox{J.S. Whitaker\unskip,\iBU}
\mbox{S.L. White\unskip,\iTENN}
\mbox{F.J. Wickens\unskip,\iRAL}
\mbox{B. Williams\unskip,\iCOLO}
\mbox{D.C. Williams\unskip,\iMIT}
\mbox{S.H. Williams\unskip,\iSLAC}
\mbox{S. Willocq\unskip,\iMASS}
\mbox{R.J. Wilson\unskip,\iCSU}
\mbox{W.J. Wisniewski\unskip,\iSLAC}
\mbox{J. L. Wittlin\unskip,\iMASS}
\mbox{M. Woods\unskip,\iSLAC}
\mbox{G.B. Word\unskip,\iVAND}
\mbox{T.R. Wright\unskip,\iWISC}
\mbox{J. Wyss\unskip,\iPADO}
\mbox{R.K. Yamamoto\unskip,\iMIT}
\mbox{J.M. Yamartino\unskip,\iMIT}
\mbox{X. Yang\unskip,\iOREG}
\mbox{J. Yashima\unskip,\iTOHO}
\mbox{S.J. Yellin\unskip,\iUCSB}
\mbox{C.C. Young\unskip,\iSLAC}
\mbox{H. Yuta\unskip,\iAOMORI}
\mbox{G. Zapalac\unskip,\iWISC}
\mbox{R.W. Zdarko\unskip,\iSLAC}
\mbox{J. Zhou\unskip.\iOREG}

\it
  \vskip \baselineskip                   % \bigskip did not work
  \centerline{(The SLD Collaboration)}   % include collaboration name
  \vskip \baselineskip        
  \baselineskip=.75\baselineskip   % shrink the interline spacing
\iADEL
  Adelphi University,
  South Avenue-   Garden City,NY 11530, \break
\iAOMORI
  Aomori University,
  2-3-1 Kohata, Aomori City, 030 Japan, \break
\iBOLO
  INFN Sezione di Bologna,
  Via Irnerio 46    I-40126 Bologna  (Italy), \break
\iBRUN
  Brunel University,
  Uxbridge, Middlesex - UB8 3PH United Kingdom, \break
\iBU
  Boston University,
  590 Commonwealth Ave. - Boston,MA 02215, \break
\iCINC
  University of Cincinnati,
  Cincinnati,OH 45221, \break
\iCOLO
  University of Colorado,
  Campus Box 390 - Boulder,CO 80309, \break
\iCOLU
  Columbia University,
  Nevis Laboratories  P.O.Box 137 - Irvington,NY 10533, \break
\iCSU
  Colorado State University,
  Ft. Collins,CO 80523, \break
\iFERR
  INFN Sezione di Ferrara,
  Via Paradiso,12 - I-44100 Ferrara (Italy), \break
\iFRAS
  Lab. Nazionali di Frascati,
  Casella Postale 13   I-00044 Frascati (Italy), \break
\iILLI
  University of Illinois,
  1110 West Green St.  Urbana,IL 61801, \break
\iLBL
  Lawrence Berkeley Laboratory,
  Dept.of Physics 50B-5211 University of California-  Berkeley,CA 94720, \break
\iLTU
  Louisiana Technical University,
  , \break
\iMASS
  University of Massachusetts,
  Amherst,MA 01003, \break
\iMISSI
  University of Mississippi,
  University,MS 38677, \break
\iMIT
  Massachusetts Institute of Technology,
  77 Massachussetts Avenue  Cambridge,MA 02139, \break
\iMOSCOW
  Moscow State University,
  Institute of Nuclear Physics  119899 Moscow  Russia, \break
\iNAGO
  Nagoya University,
  Nagoya 464 Japan, \break
\iOREG
  University of Oregon,
  Department of Physics  Eugene,OR 97403, \break
\iOXF
  Oxford University,
  Oxford, OX1 3RH, United Kingdom, \break
\iPADO
  Universita di Padova,
  Via F. Marzolo,8   I-35100 Padova (Italy), \break
\iPERU
  Universita di Perugia, Sezione INFN,
  Via A. Pascoli  I-06100 Perugia (Italy), \break
\iPISA
  INFN, Sezione di Pisa,
  Via Livornese,582/AS  Piero a Grado  I-56010 Pisa (Italy), \break
\iRAL
  Rutherford Appleton Laboratory,
  Chiton,Didcot - Oxon OX11 0QX United Kingdom, \break
\iRUTG
  Rutgers University,
  Serin Physics Labs  Piscataway,NJ 08855-0849, \break
\iSLAC
  Stanford Linear Accelerator Center,
  2575 Sand Hill Road  Menlo Park,CA 94025, \break
\iSOGA
  Sogang University,
  Ricci Hall  Seoul, Korea, \break
\iSOONG
  Soongsil University,
  Dongjakgu Sangdo 5 dong 1-1    Seoul, Korea 156-743, \break
\iTENN
  University of Tennessee,
  401 A.H. Nielsen Physics Blg.  -  Knoxville,Tennessee 37996-1200, \break
\iTOHO
  Tohoku University,
  Bubble Chamber Lab. - Aramaki - Sendai 980 (Japan), \break
\iUCSB
  U.C. Santa Barbara,
  3019 Broida Hall  Santa Barbara,CA 93106, \break
\iUCSC
  U.C. Santa Cruz,
  Santa Cruz,CA 95064, \break
\iVAND
  Vanderbilt University,
  Stevenson Center,Room 5333  P.O.Box 1807,Station B  Nashville,TN 37235,
\break
\iWASH
  University of Washington,
  Seattle,WA 98105, \break
\iWISC
  University of Wisconsin,
  1150 University Avenue  Madison,WS 53706, \break
\iYALE
  Yale University,
  5th Floor Gibbs Lab. - P.O.Box 208121 - New Haven,CT 06520-8121. \break

\rm
%
%  }   % end of address list

\end{center}

\vskip 1truecm
 
\noindent
\begin{table}
\caption{Contributions to the relative systematic error on $A_P$.}
\begin{center}
\begin{tabular}{|lc|} \hline
\qquad Error Source  &   $ \delta{A_P}/A_P$  \\ \hline
\qquad $\alpha_b$  & 5.7\%  \\
\qquad Monte Carlo statistics on $\lambda$ & 1.0\%  \\
\qquad Tag Composition  & 3.9\%  \\
\qquad Detector Modeling & 1.0\%  \\
\qquad Beam Polarization & 0.8\%  \\ \hline
\qquad Total  & 7.1\%  \\ \hline
\end{tabular}
\end{center}
\label{sys}
\end{table}

\pagebreak

\section*{Figure Captions }
 
\noindent
{\bf Figure 1}.
$P_t$-corrected vertex mass distribution for (a) highest energy jets,
(b) second-highest energy jets, and (c) lowest energy jets.
The histograms are Monte Carlo simulations; the flavor compositions 
of the simulations are indicated.
 
\vskip\baselineskip

\noindent
{\bf Figure 2}.
Numbers of significant tracks in b-tagged events 
for (a) highest energy jets, 
(b) second-highest energy jets, and (c) lowest energy jets.
The histograms are Monte Carlo simulations; the flavor compositions 
of the simulations are indicated.

\vskip\baselineskip
 
\noindent
{\bf Figure 3}.
Polar-angle distribution of the signed-thrust axis direction  with respect 
to the electron-beam direction for
(a) left-handed and (b) right-handed electron beam.
The histograms are Monte Carlo estimations of the backgrounds.

\vskip\baselineskip 

\noindent
{\bf Figure 4}.
Azimuthal-angle distribution of the signed-thrust axis direction  with respect 
to the electron-beam direction for
(a) left-handed and (b) right-handed electron beam.
The histograms are Monte Carlo estimations of the backgrounds.

\vskip\baselineskip 

\noindent
{\bf Figure 5}.
Left-right-forward-backward asymmetry in 
polar-angle distribution of the vector normal to the event plane for
(a) CP-even case, and (b) CP-odd case.
 The solid curve is the best fit to the data sample, and the dashed curves 
correspond to the 95\% C.L. limits. 

\pagebreak

\epsfysize=8in
\epsfbox{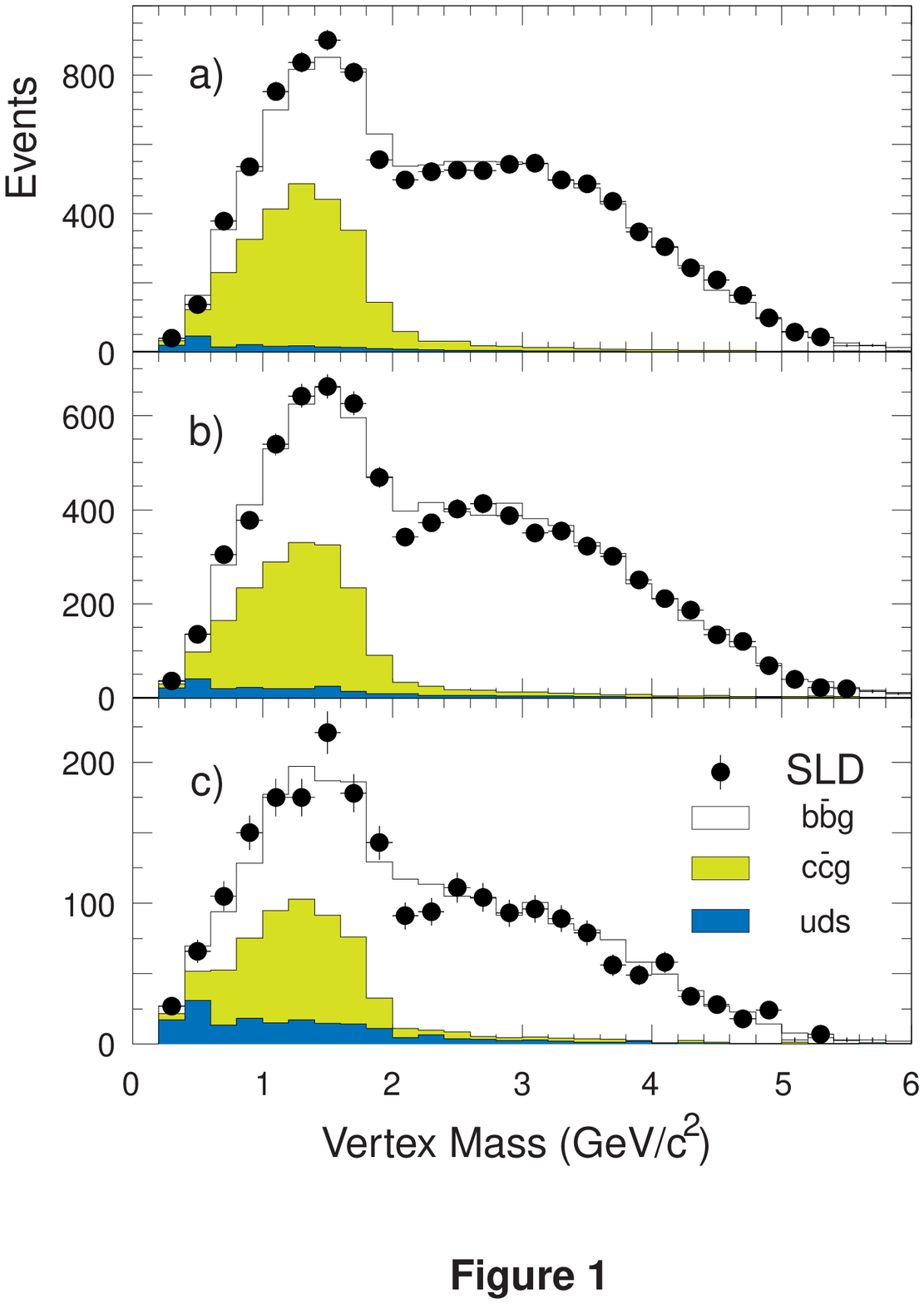}
 
\pagebreak

\epsfbox{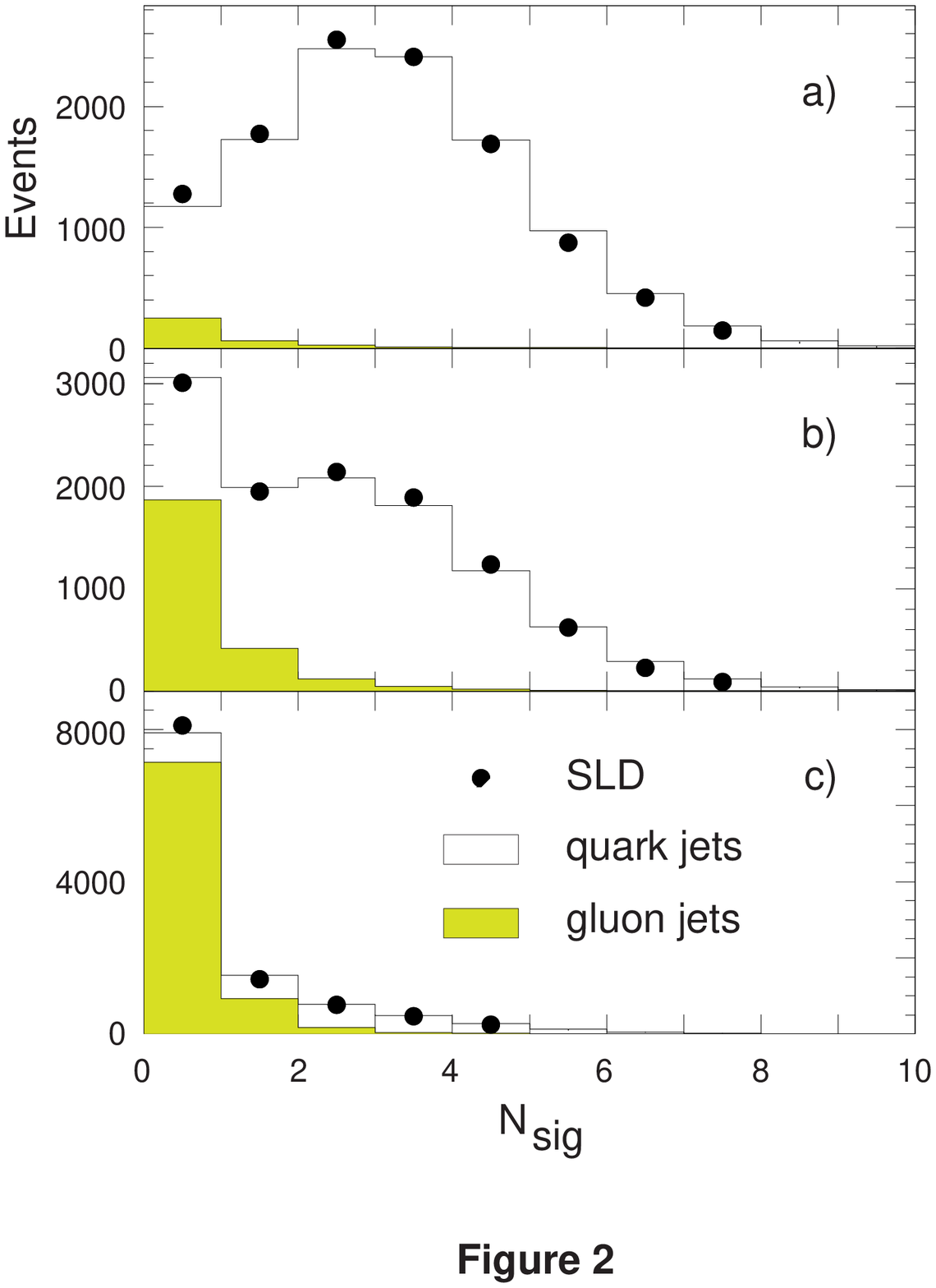}

\pagebreak

\epsfbox{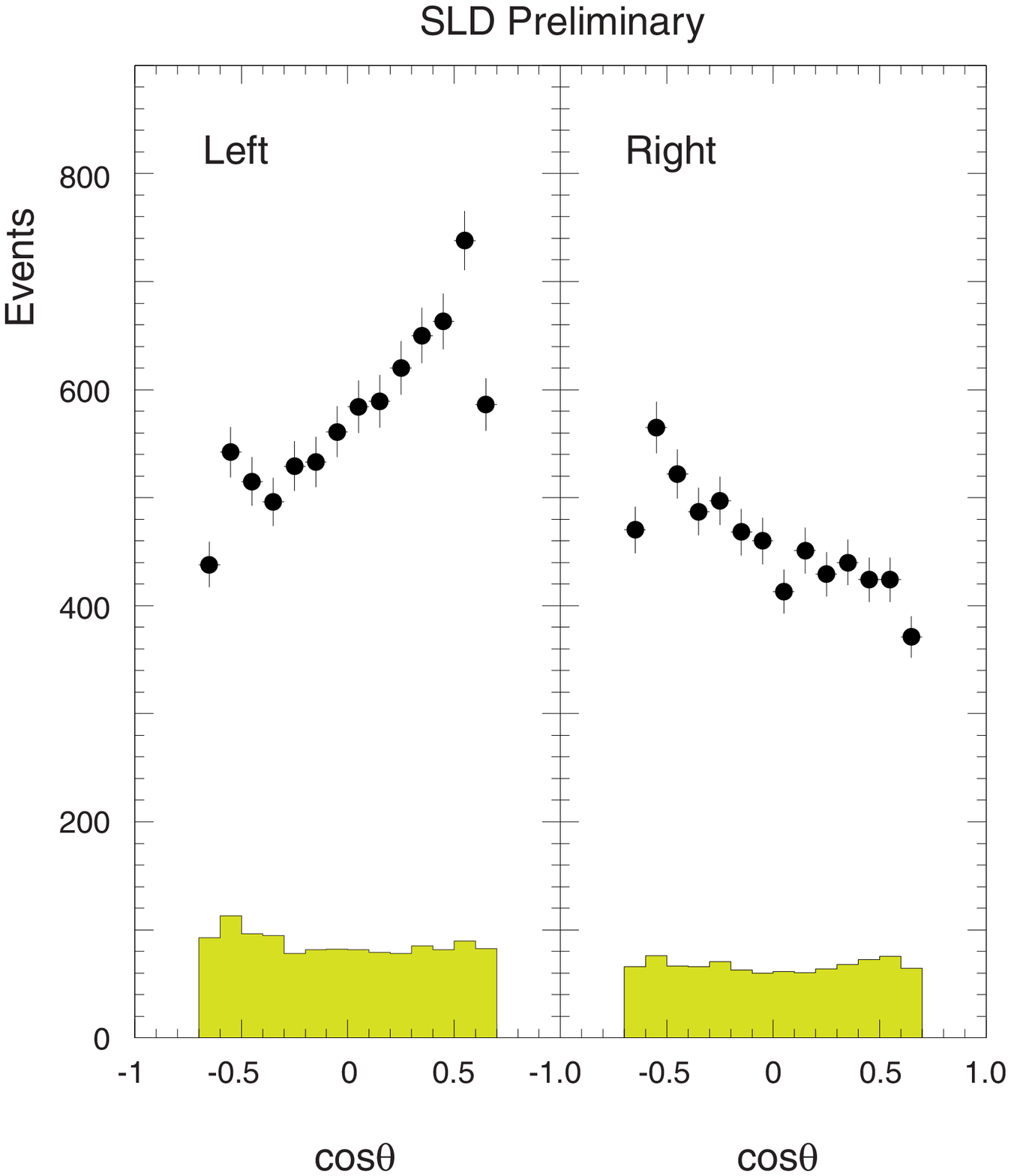}

\pagebreak

\epsfbox{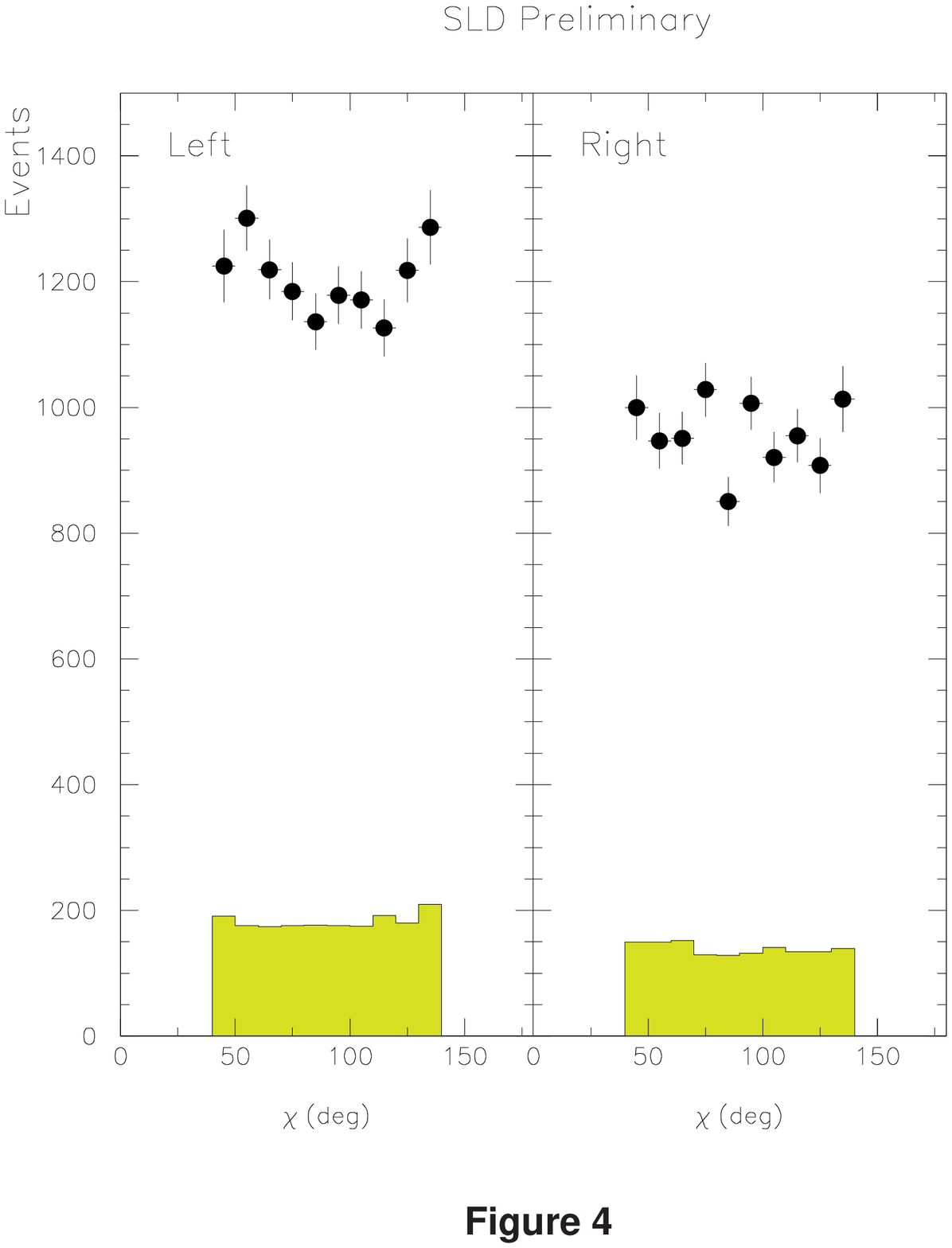}

\pagebreak

\epsfbox{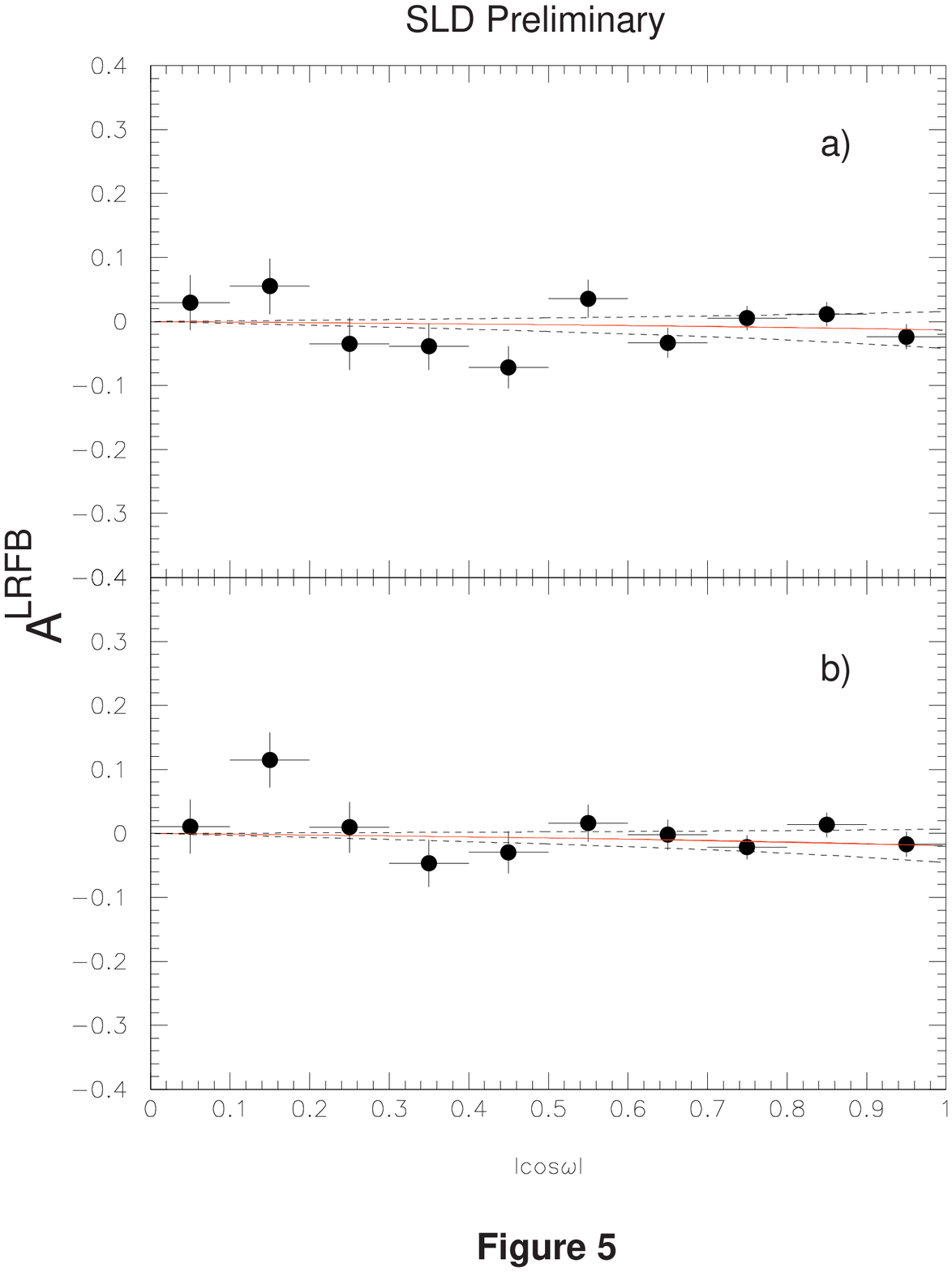}
 

\begin{thebibliography}{99}
\bibitem{miss}
The missing momentum is calculated by taking into account the measurement 
resolution of the primary and secondary vertices. We further restrict 
$M$ to be $\le 2\times M_{ch}$ to reduce contaminations from fake 
vertices in light quark events.
\end{thebibliography}
\end{document}